# A transport theory approach to percolation of liquids through porous media

D. Mostacci[a], V. Molinari, M. Premuda

*Laboratorio di Ingegneria Nucleare di Montecuccolino,*
*Alma Mater Studiorum Università di Bologna*
*via dei Colli 16, 1-40136 Bologna (Italy, EU)*

## Abstract

The percolation of a liquid through a porous material is investigated with the help of equations of the Onsager type. An expression is derived for the molecular attraction, starting from Sutherland's potential approximation to the van der Waals interaction. Then appropriate Onsager equations incorporating this molecular attraction are written from transport theory considerations, in terms of dimensionless variables. As an application, the system of self-similar equations so derived is applied to a simplified situation.

## 1. Introduction

The motion of fluids through porous media is a subject of interest in many fields of engineering and medicine [1]. It is often tackled in the framework of diffusion approximation [2]: however, this approach can only take into account density gradients, whereas temperature distribution also plays an important role in transport phenomena. The interplay of density and temperature gradients is best analyzed with the help of Onsager-type equations [3,4]. In the present work, a method is presented that is based on an Onsager equations approach to the problem of a liquid diffusing through a porous medium.

Onsager equations are, originally, a phenomenological description of non-equilibrium thermodynamics: but once they are written down in terms of gradients and coefficients, these latter still need to be determined for the equations to be useful. This can perhaps be done experimentally, or by yet other means: however, it is possible to derive them from kinetic theory, making suitable approximations on the distribution function and then making use of Boltzmann's equation [5]. In this sense, they become a rigorous result of kinetic theory, and a useful tool in those physical instances in which legitimate use can be made of them, allowance being made for the assumptions and limitations introduced in their derivation [6].

The first problem presenting itself is how to incorporate the effect of liquid state into the Boltzmann-

[a] Corresponding author - fax: +39-051-6441747; tel.: +39-051-6441711; e-mail: *domiziano.mostacci@unibo.it*

Vlasov equation that will be the starting point of the present work: this entails calculating the self consistent Vlasov field due to the intermolecular forces. Once this field is calculated, then from expanding Boltzmann-Vlasov equation in series of spherical harmonics and truncating to the first two terms, Onsager-type equations are derived for a liquid diffusing in a slab of porous medium subjected to temperature and density gradients. The coefficients of these equations are calculated considering van der Waals type interactions.

The situation considered in the present work is as follows: a slab of porous material, having breadth and length large enough compared to thickness that the problem may be viewed as one-dimensional, is in contact with a liquid having known temperature $T_0$ and density $n_0$ on one side, and $T_L$ and $n_L$ on the other side, see figure 1. The liquid diffuses through the porous material under the effect of density and temperature gradients.

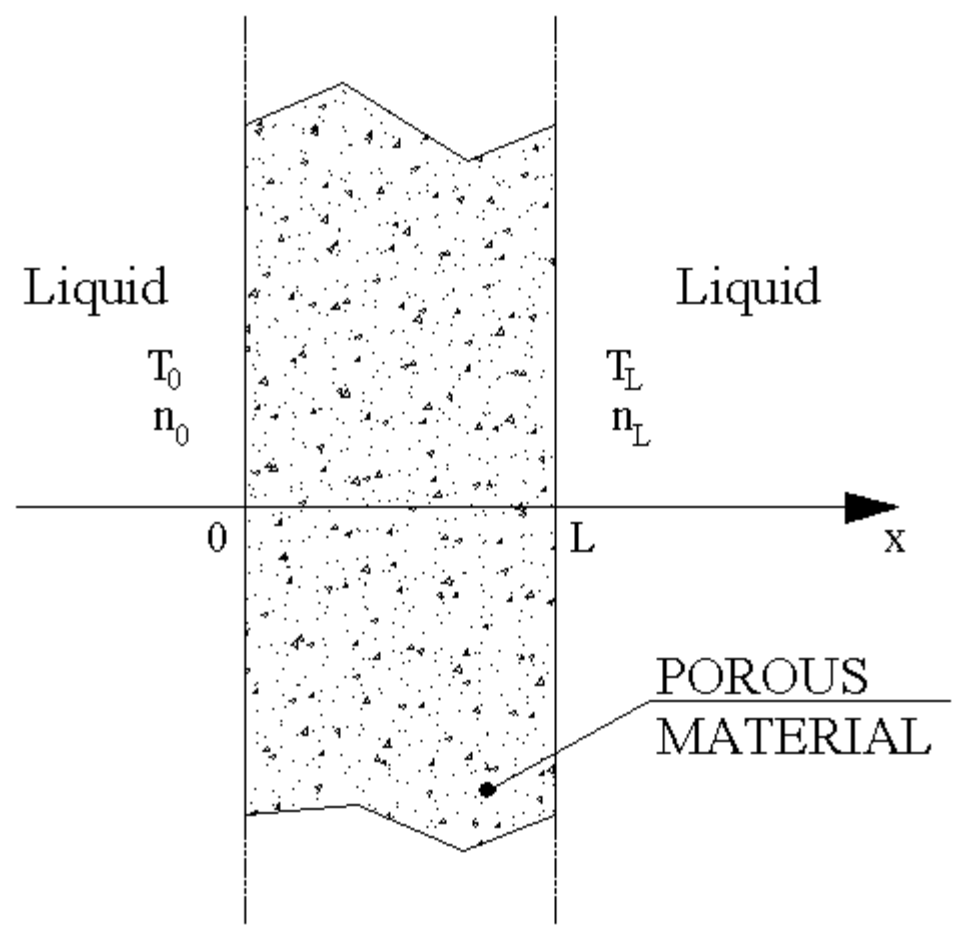

**Figure 1**: geometry of the problem

Onsager equations are then solved to obtain temperature, density and pressure profiles inside the slab, and from these latter the distribution function for the liquid molecules.

## 2. Self consistent Vlasov field from van der Waals interaction

The percolating liquid molecules undergo mutual interactions as well as interactions with the medium. This latter effect will be discussed in the next section. Regarding the interactions between molecules, two categories can be recognized: binary short range collisions and long range interactions. Collisions are responsible for forcing the system toward a maxwellian equilibrium distribution function, and will not be studied further in the present work: it will be assumed that the isotropic part of the distribution function is not far from having a maxwellian distribution. It may be recalled that in a perfect gas no other interaction would be present. Long range interactions are by nature multi-body, and are best investigated introducing a Vlasov self consistent field. The forces responsible for the long range interactions between non-polar molecules are generally assumed to be only function of the separation r. In the present work the intermolecular force will be modeled with the Sutherland potential $\varphi(r)$, defined as follows [7, 8, 9]:

$$\varphi(r) = \begin{cases} -4\varepsilon\left(\dfrac{\sigma}{r}\right)^{\alpha} & \text{for } r \geq \sigma \\ \infty & \text{for } r < \sigma \end{cases} \tag{1}$$

representing rigid spheres of diameter $\sigma$ attracting each other with an inverse power $\alpha$ law. For $\alpha = 6$ this potential represents the van der Waals attraction. The parameters $\varepsilon$ and $\sigma$ can be found tabulated for various liquids in many sources, e.g., [7, 8].

From the model discussed above, the self consistent field has been derived for $\alpha = 6$ for the stationary 1-D case. Details of the derivation can be found in Appendix. The end result is the following expression for the Vlasov self consistent force $F_L$:

$$F_L = 8\pi\varepsilon\sigma^6\left\{-\int_{-\infty}^{z-\sigma}\frac{n(\zeta)}{(z-\zeta)^5}d\zeta + \int_{z+\sigma}^{+\infty}\frac{n(\zeta)}{(\zeta-z)^5}d\zeta\right\} \tag{2}$$

where $n(z)$ is the (unknown) molecule number density at location z. If the density $n(z)$ is constant at all location, then the overall force vanishes. Again, the above expression has been obtained considering only the effect of collective motion, neglecting correlations. Therefore it is particularly suited for investigating organized collective motions, such as waves, or particular effects in finite media, such as surface tension [9], or yet the diffusion and heat flux generated by density and temperature gradients.

If the density variation is mild, $n(z)$ can be expanded in Taylor series, and retaining only the first few terms an approximate value for $F_L$ obtained. The procedure is shown in the appendix, the end result being

$$F_L(z) \cong \Lambda\frac{d\,n(z)}{dz} + \Lambda_3\frac{d^3n(z)}{dz^3} \tag{3}$$

where the coefficients are given by

$$\Lambda = \frac{16\pi}{3}\varepsilon\sigma^3 \qquad\qquad \Lambda_3 = \frac{16\pi}{6}\varepsilon\sigma^5 \tag{4}$$

## 3. Percolation model and Onsager equations

As discussed above, in references [5,6,10] Onsager-type transport equations were derived from Boltzmann equation; transport coefficients can then be determined given the appropriate cross sections. The starting point is the stationary form of the Boltzmann equation [5,6]

$$\mathbf{v}\cdot\frac{\partial f}{\partial \mathbf{r}} + \frac{\mathbf{F}_L}{m}\cdot\frac{\partial f}{\partial \mathbf{v}} = \left(\frac{\partial f}{\partial t}\right)_{coll} \tag{5}$$

Here f is the distribution function of the particles considered. Expanding f in spherical harmonics in Ω and truncating to the first two terms, the following expression for f is obtained:

$$f(\mathbf{r}, v, \boldsymbol{\Omega}) = \frac{1}{4\pi}\left[f_0(\mathbf{r}, v) + 3\boldsymbol{\Omega} \bullet \mathbf{f}_1(\mathbf{r}, v)\right] \tag{6}$$

where $f_0$ and $\mathbf{f}_1$ are the moments of order zero and one of the distribution function. In this approximation, upon solving Boltzmann equation for $\mathbf{f}_1$ in the present 1-D case, an expression for $f_1$ as a function of $f_0$ is obtained [5,11]

$$v\frac{\partial f_0}{\partial z} + \frac{F}{m}\frac{\partial f_0}{\partial v} = -3\nu f_1(z, v) \tag{7}$$

where ν is the collision frequency.

As is customary in transport theory, interactions between the molecules of the liquid and the diffusing medium will be described through cross sections. In the following the so called rigid sphere model will be assumed for the cross sections, and the medium will be thought of as composed of grains of varying size, as depicted schematically in figure 2, all interacting with liquid molecules as rigid spheres [12].

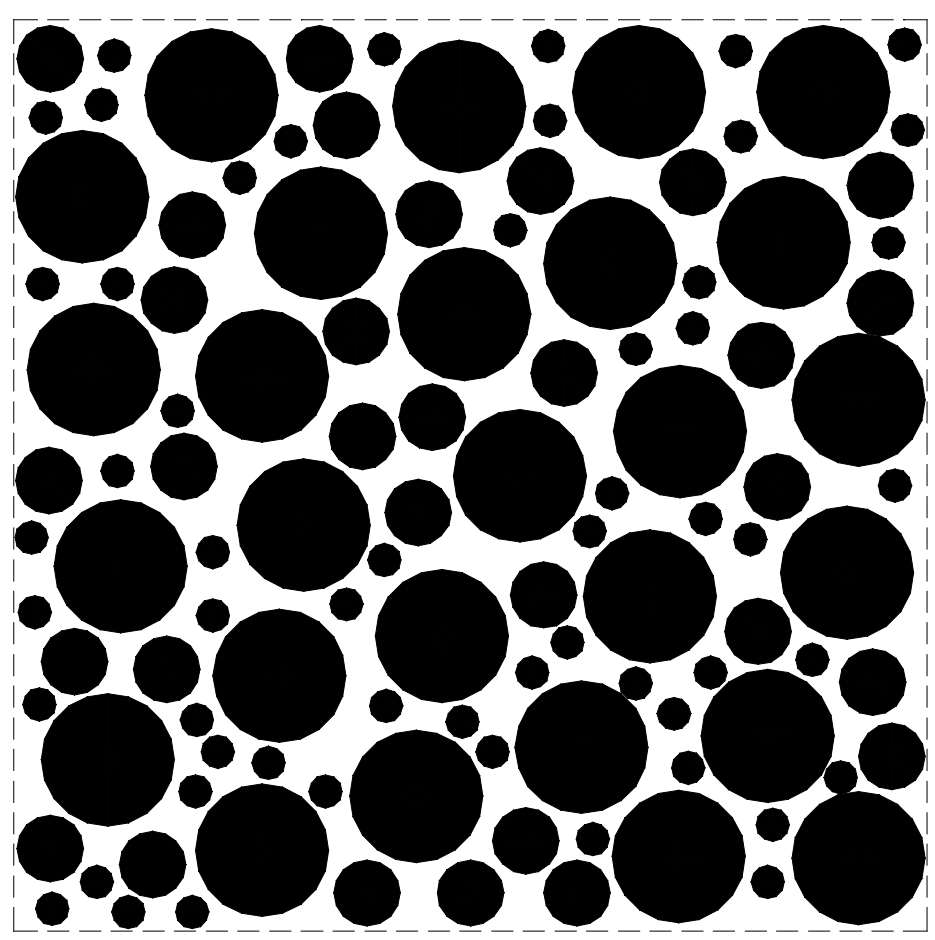

**Figure 2**: porous medium model

Also, grains will be very massive compared to the molecules, and fixed in the laboratory frame, so that scattering is isotropic in the laboratory reference frame. Under such conditions, the transport cross section is independent of molecule velocity. Different sizes will entail different microscopic cross sections for the grains, yet they can all be compounded into one macroscopic cross section $\Sigma_0$, independent of velocity, with the usual methods of kinetic theory. Under these conditions, the collision frequency is given by

$$\nu(v) = v\Sigma_0 \tag{8}$$

As will appear clearly in the following, the exact value of this cross section is not needed to solve the problem as was proposed, its being a constant suffices to determine the requested profiles.

The force term F accounts for any type of force, external and self-consistent: in the present case the intermolecular force derived in the previous section will be considered alone, and in keeping with the P1 approximation of (6) only the first term will be retained. Then the following generalized form of Fick's law can be wrtitten

$$f_1(z,v) = -\frac{v}{3\nu}\frac{\partial f_0}{\partial z} - \frac{1}{3\nu}\frac{F_L}{m}\frac{\partial f_0}{\partial v} \tag{9}$$

Also, the distribution function can now be calculated as

$$f(z,v,\mu) = \frac{1}{4\pi}\left[f_0(z,v) - \mu\left(\frac{v}{\nu}\frac{\partial f_0}{\partial z} + \frac{F_L}{m\nu}\frac{\partial f_0}{\partial v}\right)\right] \tag{10}$$

The particle current and energy flux are defined, as usual, as

$$J = \frac{4\pi}{3}\int_0^\infty f_1 v^5 dv\ ; \qquad H = \frac{2\pi m}{3}\int_0^\infty f_1 v^5 dv \tag{11}$$

yielding

$$J = -D_L\frac{\partial n}{\partial x} - \alpha\frac{\partial T}{\partial x} \tag{12-a}$$

$$H = -\gamma_L\frac{\partial n}{\partial x} - \kappa\frac{\partial T}{\partial x} \tag{12-b}$$

with coefficients defined as

$$D_L = \frac{4\pi}{3}\frac{1}{n}\left(1 - \frac{n\Lambda}{T}\right)\int_0^\infty \frac{v^4 f_0}{\nu}dv \tag{13-a}$$

$$\alpha = \frac{4\pi}{3}\frac{1}{T}\int_0^\infty \frac{v^4 f_0}{\nu}\left(\frac{mv^2}{2T} - \frac{3}{2}\right)dv \tag{13-b}$$

$$\gamma_L = \frac{2\pi}{3}\frac{m}{n}\left(1 - \frac{n\Lambda}{T}\right)\int_0^\infty \frac{v^6 f_0}{\nu}dv \tag{13-c}$$

$$\kappa = \frac{2\pi}{3}\frac{m}{T}\int_0^\infty \frac{v^6 f_0}{\nu}\left(\frac{mv^2}{2T} - \frac{3}{2}\right)dv \tag{13-d}$$

If the isotropic part of the distribution function can be approximated as a maxwellian, the above coefficients are calculated as

$$D_L = D_c\left(1 - \frac{n\Lambda}{T}\right)T^{\frac{1}{2}} \tag{14a}$$

$$\alpha = D_c \frac{1}{2} n T^{-\frac{1}{2}} \tag{14b}$$

$$\gamma_L = D_c 2\left(1 - \frac{n\Lambda}{T}\right) T^{\frac{3}{2}} \tag{14c}$$

$$\kappa = D_c 3 n T^{\frac{1}{2}} \tag{14d}$$

$$D_c = \frac{2}{3\Sigma_0}\sqrt{\frac{2}{m\pi}} \tag{15}$$

## 4. Solution of the Onsager equations

The problem can be recast into a self-similar form by defining the following dimensionless quantities:

$$\tilde{x} = \frac{x}{L}; \qquad \tilde{T} = \frac{T}{T_0}; \qquad \tilde{n} = \frac{n}{n_0}; \qquad \varsigma = \frac{\Lambda n_0}{T_0};$$
$$\tilde{H} = \frac{H}{H_0}; \qquad \tilde{J} = \frac{J}{J_0}; \qquad H_0 = \frac{2 D_c n_0 T_0^{\frac{3}{2}}}{L} \qquad J_0 = \frac{H_0}{2T_0} \tag{16}$$

The quantity $\varsigma$ represents a ratio of the cohesive energy of the liquid to the energy of thermal agitation. A quick calculation shows how $H_0$ is the heat flow that would be obtained for vanishing temperature $T_L$. Introducing the coefficients (14) into the (12) the following system of two equations is obtained:

$$\tilde{J} = -\left[\left(1 - \varsigma\frac{\tilde{n}}{\tilde{T}}\right)\frac{1}{\tilde{n}}\frac{d\tilde{n}}{d\tilde{x}} + \frac{1}{2}\frac{1}{\tilde{T}}\frac{d\tilde{T}}{d\tilde{x}}\right]\tilde{n}\,\tilde{T}^{\frac{1}{2}} \tag{17a}$$

$$\tilde{H} = -\left[\left(1 - \varsigma\frac{\tilde{n}}{\tilde{T}}\right)\frac{1}{\tilde{n}}\frac{d\tilde{n}}{d\tilde{x}} + \frac{3}{2}\frac{1}{\tilde{T}}\frac{d\tilde{T}}{d\tilde{x}}\right]\tilde{n}\,\tilde{T}^{\frac{3}{2}} \tag{17b}$$

Calling $\tilde{R}$ the ratio of current to energy flow:

$$\tilde{R} = \frac{\tilde{J}}{\tilde{H}} \tag{18}$$

(which can be seen to satisfy $\tilde{R} \in \left(\frac{1}{3}, 1\right)$ upon considering that it is the ratio between (17a) and (17b)), the following two equations can be obtained:

$$\frac{d\tilde{T}}{d\tilde{n}} = 2\left(\varsigma - \frac{\tilde{T}}{\tilde{n}}\right)\frac{1 - \tilde{R}\tilde{T}}{1 - 3\tilde{R}\tilde{T}} \tag{19}$$

$$\frac{d\tilde{T}}{d\tilde{x}} = -\tilde{H}\,\frac{1 - \tilde{R}\tilde{T}}{\tilde{n}\sqrt{\tilde{T}}} \tag{20}$$

In the case of water at room temperature $\varsigma$ is found to be of order $10^{-2}$, so it can be neglected in calcualting the temperature profile. Then (19) simplifies to

$$\frac{d\tilde{T}}{d\tilde{n}} = -2\frac{\tilde{T}}{\tilde{n}}\frac{1-\tilde{R}\tilde{T}}{1-3\tilde{R}\tilde{T}} \tag{21}$$

yielding a relation between temperature and density:

$$\tilde{n} = \frac{1}{\sqrt{\tilde{T}}}\frac{1-\tilde{R}}{1-\tilde{R}\tilde{T}} \tag{22}$$

and then (20) can be solved to yield

$$\tilde{T} = \frac{1-\tilde{H}\,\tilde{x}}{1-\tilde{R}\tilde{H}\,\tilde{x}} \tag{23}$$

It can be seen how it must be $\tilde{H} < 1$, i.e., $H < H_0$, in keeping with the remark following the definition of $H_0$ (16).

The density can now be calculated from the (19) rewritten as

$$\frac{d\tilde{n}}{d\tilde{T}} = \frac{1}{2}\frac{\tilde{n}}{\tilde{n}\varsigma - \tilde{T}}\frac{1-3\tilde{R}\tilde{T}}{1-\tilde{R}\tilde{T}} \tag{24}$$

To solve the above equation, keeping in mind that $\varsigma$ is quite small and, being a liquid, $\tilde{n}$ cannot differ much from its initial value of one, the following simplified equation will be considered:

$$\frac{d\tilde{n}}{d\tilde{T}} = \frac{1}{2}\frac{\tilde{n}}{\bar{n}\varsigma - \tilde{T}}\frac{1-3\tilde{R}\tilde{T}}{1-\tilde{R}\tilde{T}} \tag{25}$$

where $\bar{n}$ is some value of the dimensionless density. This has solution

$$\tilde{n} = \left[\frac{1}{\sqrt{\tilde{T}}}\frac{1-\tilde{R}}{1-\tilde{R}\tilde{T}}\right] \times \sqrt{\tilde{T}\frac{1-\bar{n}\varsigma}{\tilde{T}-\bar{n}\varsigma}\left[\frac{1-\tilde{R}}{1-\tilde{R}\tilde{T}}\frac{\tilde{T}-\bar{n}\varsigma}{1-\bar{n}\varsigma}\right]^{\tilde{R}\bar{n}\varsigma}} \tag{26}$$

In the graphs reported $\bar{n} = n(0)$ has been chosen.

## 5. Conclusions

The results of the method, applied to the simplified test problem described previously, can be seen in the following figures 3-6. As independent variables, i.e., the parameters specified externally, the quantities $\tilde{T}_1 = \tilde{T}(1)$ and $\tilde{J}$ have been chosen. This corresponds to specifying the boundary temperatures and the flow of liquid.

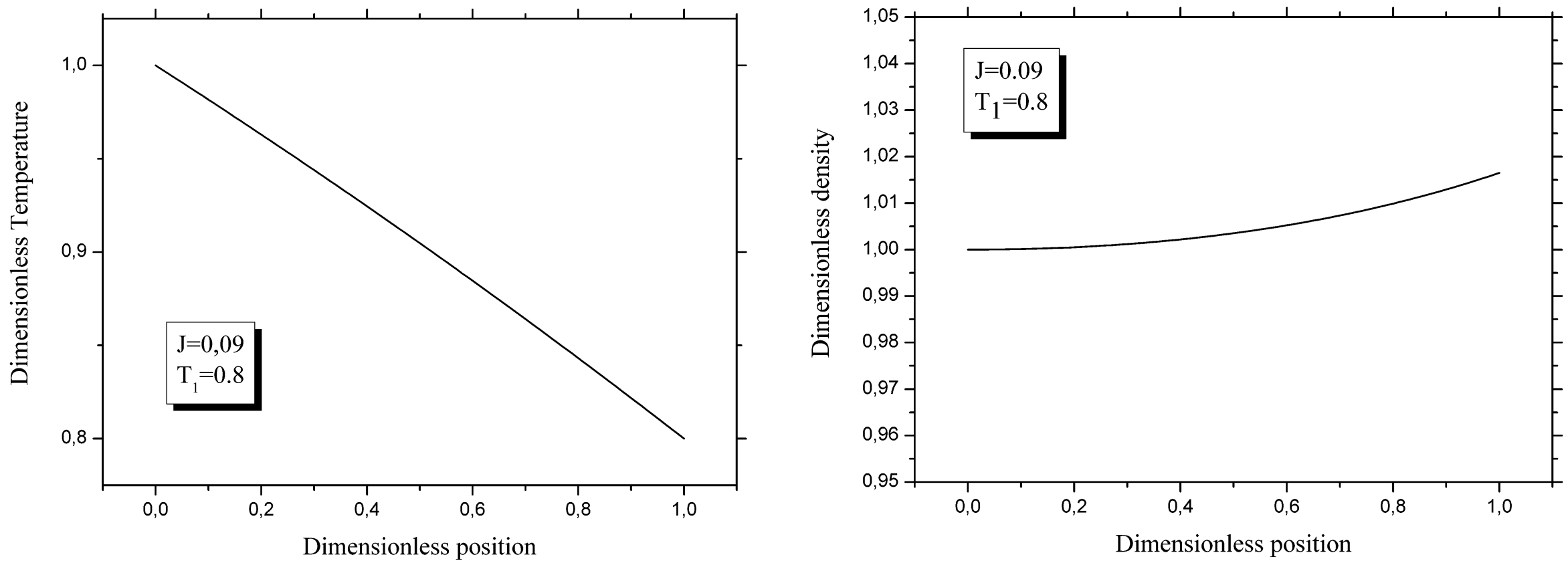

**Figure 3**: temperature and density profiles for $\tilde{T}_1 = 0.8$ and $\tilde{J} = 0.09$

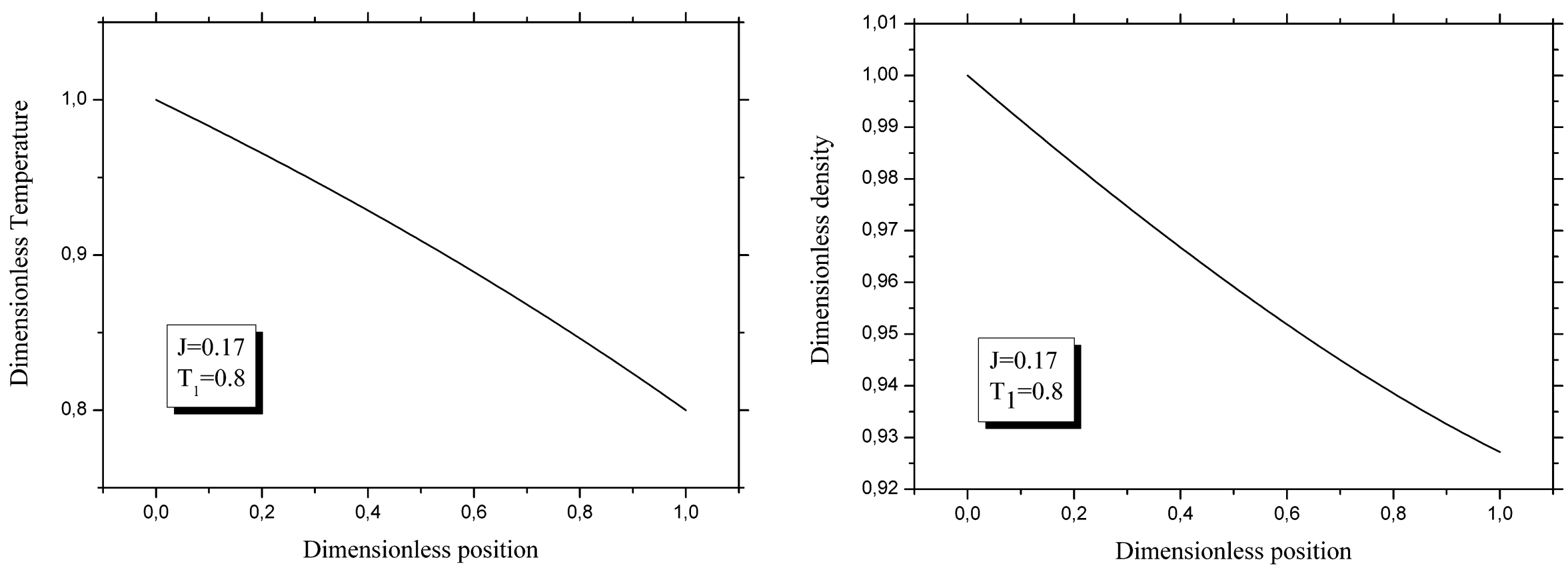

**Figure 4**: temperature and density profiles for $\tilde{T}_1 = 0.8$ and $\tilde{J} = 0.17$

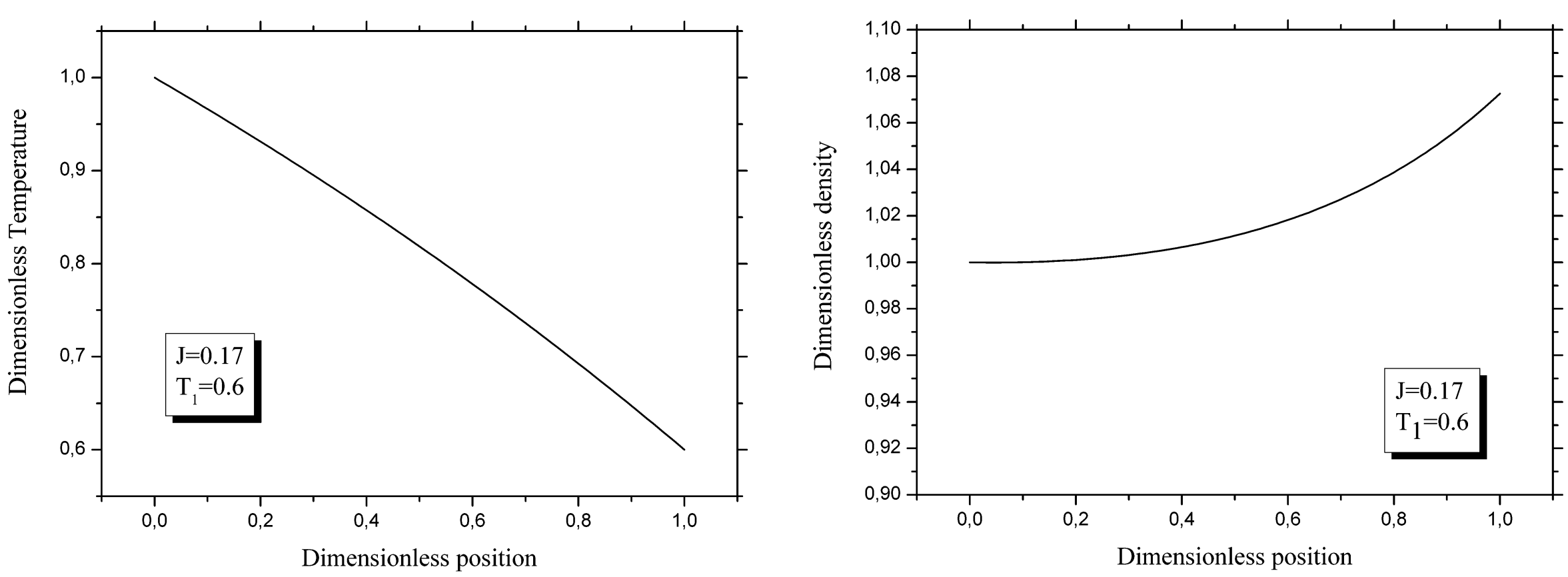

**Figure 5**: temperature and density profiles for $\tilde{T}_1 = 0.6$ and $\tilde{J} = 0.17$

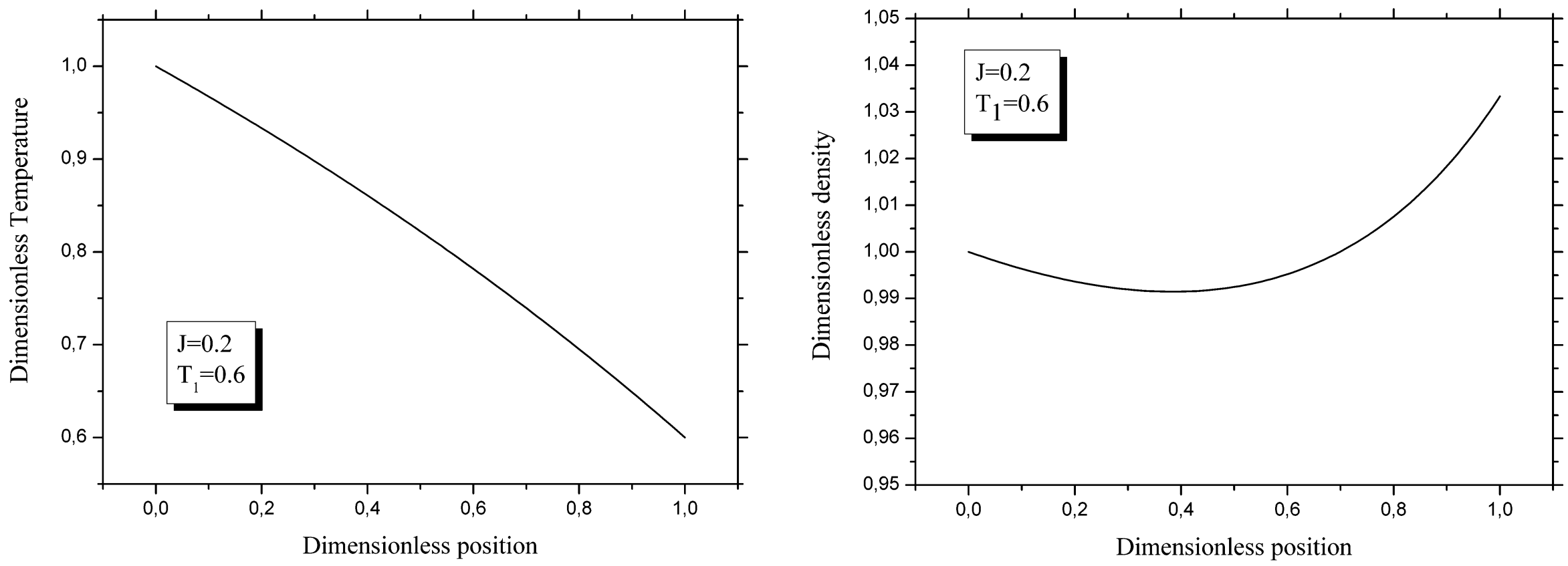

**Figure 6**: temperature and density profiles for $\tilde{T}_1 = 0.6$ and $\tilde{J} = 0.20$

As can be seen the general shape of the temperature profile does not exhibit a strong variation in the different cases. Density profiles, however, are influenced more markedly by the temperature difference and the intensity of the liquid flow. It is to be underlined, though, that the choice of $\tilde{J}$ is somewhat arbitrary, and that the density profile adjusts itself to this choice.

One last point to recall is that from the knowledge of temperature and density, the distribution function at every point can be recovered with the help of (10).

## Appendix

It is desired to calculate the self consistent field from the molecular attraction. To this end a molecule located in (0,0,z) will be considered, see figure A.1, and the force from the whole surrounding liquid calculated.

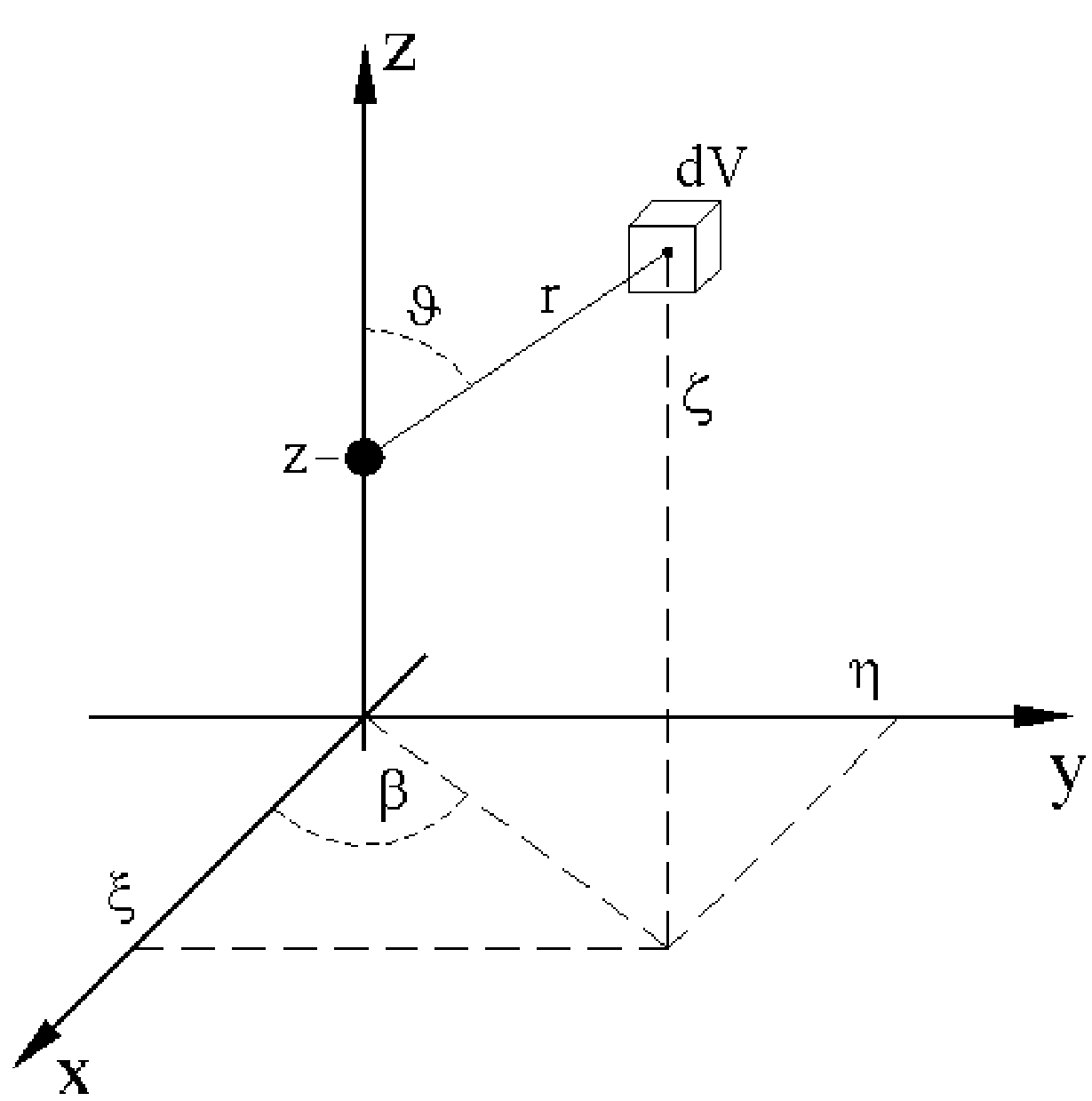

**Figure A1**: geometry of the setting

To simplify the problem, a system possessing slab symmetry will be considered, in which density depends only on the z-coordinate. Consider then an elementary volume dV at a location defined by the coordinates $(r,\vartheta,\beta)$ in a spherical reference system centered in the molecule of interest and with the polar axis along the z-direction.

The potential in $(0,0,z)$ due to a molecule in $(r,\vartheta,\beta)$ is given by

$$\varphi(r) = -\frac{G}{r^{\alpha}} \tag{A.1}$$

and hence the force on the molecule in $(0,0,z)$

$$\mathbf{F} = -\nabla\varphi(r) = \frac{\alpha G}{r^{\alpha+1}}\hat{\mathbf{r}} \tag{A.2}$$

In cartesian coordinates this becomes

$$\mathbf{F}_1 = F_{1x}\hat{\mathbf{x}} + F_{1y}\hat{\mathbf{y}} + F_{1z}\hat{\mathbf{z}} \tag{A.3}$$

and calling $(\xi,\eta,\zeta)$ the cartesian coordinates of the center of the volume dV, the distance r can be calculated as

$$r^{\alpha+1} = \left[\xi^2 + \eta^2 + (\zeta - z)^2\right]^{\frac{\alpha+1}{2}} \tag{A.4}$$

and the cartesian components of the force

$$F_{1x} = \frac{\alpha G\, n(\zeta) d\xi d\eta d\zeta}{r^{\alpha+1}} \sin\vartheta \cos\beta \tag{A.5}$$

$$F_{1y} = \frac{\alpha G\, n(\zeta) d\xi d\eta d\zeta}{r^{\alpha+1}} \sin\vartheta \sin\beta \tag{A.6}$$

$$F_{1z} = \frac{\alpha G\, n(\zeta) d\xi d\eta d\zeta}{r^{\alpha+1}} \cos\vartheta \tag{A.7}$$

Considering that the sines and cosines in the above equations can be expressed as

$$\cos\vartheta = \frac{\zeta - z}{\sqrt{\xi^2 + \eta^2 + (\zeta - z)^2}} \tag{A.8}$$

$$\sin\vartheta = \frac{\sqrt{\xi^2 + \eta^2}}{\sqrt{\xi^2 + \eta^2 + (\zeta - z)^2}} \tag{A.9}$$

$$\cos\beta = \frac{\xi}{\sqrt{\xi^2 + \eta^2}} \tag{A.10}$$

$$\sin\beta = \frac{\eta}{\sqrt{\xi^2 + \eta^2}} \tag{A.11}$$

the cartesian components of the force may be rewritten as

$$F_{1x} = \alpha G\, n(\zeta) d\xi d\eta d\zeta \frac{\xi}{\left[\xi^2 + \eta^2 + (\zeta - z)^2\right]^{\frac{\alpha+2}{2}}} \tag{A.12}$$

$$F_{1y} = \alpha G\, n(\zeta) d\xi d\eta d\zeta \frac{\eta}{\left[\xi^2 + \eta^2 + (\zeta - z)^2\right]^{\frac{\alpha+2}{2}}} \tag{A.13}$$

$$F_{1z} = \alpha G\, n(\zeta) d\xi d\eta d\zeta \frac{\zeta - z}{\left[\xi^2 + \eta^2 + (\zeta - z)^2\right]^{\frac{\alpha+2}{2}}} \tag{A.14}$$

To obtain the overall force on the reference molecule, integration over the whole volume is performed. It can be seen readily that

$$\int_{-\infty}^{+\infty}\int_{-\infty}^{+\infty} F_{1x}\, d\xi\, d\eta = \int_{-\infty}^{+\infty}\int_{-\infty}^{+\infty} F_{1y}\, d\xi\, d\eta = 0 \tag{A.15}$$

so that there are no x and y components to the force. As for the z component

$$\int_{-\infty}^{+\infty}\int_{-\infty}^{+\infty} F_{1z}\, d\xi\, d\eta = \alpha G\, n(\zeta) d\zeta (\zeta - z) \int_{-\infty}^{+\infty}\int_{-\infty}^{+\infty} \frac{1}{\left[\xi^2 + \eta^2 + (\zeta - z)^2\right]^{\frac{\alpha+2}{2}}} d\xi\, d\eta \tag{A.16}$$

Considering Sutherland's potential in (1): $\alpha = 6$, there is a minimum approach distance $\sigma$ and the constant in (A.1) becomes $G = 4\varepsilon\sigma^6$

$$F_L = 8\pi\varepsilon\sigma^6 \left\{ -\int_{-\infty}^{z-\sigma} \frac{n(\zeta)}{(z-\zeta)^5} d\zeta + \int_{z+\sigma}^{+\infty} \frac{n(\zeta)}{(\zeta - z)^5} d\zeta \right\} \tag{A.17}$$

If the density variation is mild, $n(z)$ can be expanded in Taylor series retaining only the first few terms

$$n(\zeta) = n(z) + \frac{dn(z)}{dz}(\zeta - z) + \frac{d^2 n(z)}{dz^2}\frac{(\zeta - z)^2}{2} + \frac{d^3 n(z)}{dz^3}\frac{(\zeta - z)^3}{3!} + \frac{d^3 n(z)}{dz^3}\frac{(\zeta - z)^4}{4!} + O\left[(\zeta - z)^5\right] \tag{A.18}$$

Neglecting terms of order 5 and higher, and substituting into eq. (A.17), after some algebra the following equation is obtained:

$$F_L(z) \cong \Lambda \frac{d\,n(z)}{dz} + \Lambda_3 \frac{d^3 n(z)}{dz^3} \tag{A.19}$$

where the coefficients are given by

$$\Lambda = \frac{16\pi}{3}\varepsilon\sigma^3 \qquad \Lambda_3 = \frac{16\pi}{6}\varepsilon\sigma^5 \tag{A.20}$$